\documentclass{article}
\usepackage{arxiv}

\usepackage[utf8]{inputenc} % allow utf-8 input
\usepackage[T1]{fontenc}    % use 8-bit T1 fonts
\usepackage{silence}
\usepackage{hyperref}       % hyperlinks
\usepackage{url}            % simple URL typesetting
\usepackage{booktabs}       % professional-quality tables
\usepackage{amsfonts}       % blackboard math symbols
\usepackage{nicefrac}       % compact symbols for 1/2, etc.
\usepackage{microtype}      % microtypography
\usepackage{lipsum}		% Can be removed after putting your text content
\newif\iffast
\fastfalse
\iffast
\usepackage[draft]{graphicx}
\else
\usepackage{graphicx}
\fi
\usepackage{amsmath}
\usepackage{amssymb}
\usepackage{longtable}
\usepackage{multirow}
\usepackage{array}
\usepackage{calc}
\usepackage{listings}
\usepackage[section]{placeins} % keep floats from drifting past section boundaries

\title{mViSE: A Visual Search Engine for Analyzing Multiplex IHC Brain Tissue Images}

\author{
\normalfont
Liqiang Huang\textsuperscript{1}\and
Rachel W. Mills\textsuperscript{1}\and
Saikiran Mandula\textsuperscript{1}\and
Lin Bai\textsuperscript{1}\and
Mahtab Jeyhani\textsuperscript{1}\and
John Redell\textsuperscript{2}\and
Hien Van Nguyen\textsuperscript{1}\and
Saurabh Prasad\textsuperscript{1}\and
Dragan Maric\textsuperscript{3}\and
Badrinath Roysam\textsuperscript{1} \\
\textsuperscript{1}Cullen College of Engineering, University of Houston\\ Houston, Texas 77204, USA \\
\textsuperscript{2}The University of Texas McGovern Medical School, Houston\\ Texas 77030, USA \\
\textsuperscript{3}National Institute of Neurological Disorders and Stroke\\ Bethesda, Maryland 20892, USA
}

\renewcommand{\shorttitle}{mViSE: A Visual Search Engine for Analyzing Multiplex IHC Brain Tissue Images}

\date{} % Remove date

\hypersetup{
pdftitle={mViSE: A Visual Search Engine for Analyzing Multiplex IHC Brain Tissue Images},
pdfsubject={q-bio.NC, q-bio.QM},
pdfauthor={Liqiang Huang, Rachel W. Mills, Saikiran Mandula, Lin Bai, Mahtab Jeyhani, John Redell, Hien Van Nguyen, Saurabh Prasad, Dragan Maric, Badrinath Roysam},
pdfkeywords={Visual Search Engine, Multiplex IHC, Brain Tissue Images, mViSE},
}

\begin{document}
\maketitle
\date{}
\begin{abstract}
Whole-slide multiplex imaging of brain tissue generates massive information-dense images that are challenging to analyze and require custom software. We present an alternative query-driven programming-free strategy using a multiplex visual search engine (mViSE) that learns the multifaceted brain tissue chemoarchitecture, cytoarchitecture, and myeloarchitecture. Our divide-and-conquer strategy organizes the data into panels of related molecular markers and uses self-supervised learning to train a multiplex encoder for each panel with explicit visual confirmation of successful learning. Multiple panels can be combined to process visual queries for retrieving similar communities of individual cells or multicellular niches using information-theoretic methods. The retrievals can be used for diverse purposes including tissue exploration, delineating brain regions and cortical cell layers, profiling and comparing brain regions without computer programming. We validated mViSE's ability to retrieve single cells, proximal cell pairs, tissue patches, delineate cortical layers, brain regions and sub-regions. mViSE is provided as an open-source QuPath plug-in.
\end{abstract}


\section{Introduction}
Advances in whole-slide high-resolution multispectral imaging and cyclic protocols for immunohistochemical staining are enabling the collection of large, complex, and information-dense images of whole brain slices (MP-WSI) that capture multiple cell types, cell states, tissue arrangements, and biological processes simultaneously in their cyto-histological context, e.g., the intended and side effects of a drug candidate\textsuperscript{1}. Multiplex images are challenging to analyze due to their complexity, the need for custom interpretations for each channel, complex process for establishing cell types/sub-types and phenotypic status, and large data size. Mammalian Brain tissue is especially challenging due to its scale and cell-molecular-spatial complexity\textsuperscript{2-6}, containing a heterogeneous mix of neuronal\textsuperscript{7}, glial\textsuperscript{8}, and vascular cell types\textsuperscript{9}, organized in elaborate region-specific cellular arrangements (cytoarchitectures)\textsuperscript{10}, protein distribution patterns (chemoarchitecture)\textsuperscript{6}, wiring patterns (myeloarchitecture)\textsuperscript{6}, and cyto-vascular (niche) patterns\textsuperscript{11}.

We earlier presented cyclic multiplex immunofluorescence protocols and computational methods for systematic and comprehensive profiling of brain tissue\textsuperscript{12}. The list of markers included multiple identifiers of all major cell types resident to the normal rat brain: (1) ubiquitous nucleus markers DAPI and histones, (2) neuronal myeloarchitectonic markers NeuN, microtubule-associated protein (MAP2), neurofilament-H (NF-H), neurofilament-M (NF-M) and synaptophysin, (3) neuronal phenotyping markers Tbr1, Eomes, glutaminase (GLUT), tyrosine hydroxylase (TH), choline acetyltransferase (ChAT), parvalbumin (PARVALB), GAD67, calbindin and calretinin, (4) astrocyte differentiation markers glial fibrillary acidic protein (GFAP), aquaporin-4, S100$\beta$ and GLAST, (5) oligodendrocyte markers Olig2, CNPase and myelin basic protein (MBP), vascular associated markers endogenous GFP, rat endothelial cell antigen (RECA1), smooth muscle actin (SMA) and tomato lectin (T-lectin), rat blood brain barrier (BBB) antigen and PDGFR a marker of pericytes, (6) microglial marker IBA1, (7) immature stem/progenitor cell markers Sox2, nestin, vimentin and doublecortin (8) actively proliferating cells PCNA and (9) actively apoptotic cells cleaved caspase-3 (CC-3). The multiplex image analysis was implemented as a pipeline of steps, starting with image pre-processing, cell detection, cell segmentation, identification of cell types and phenotypic states based on protein expression patterns, and aggregation of the histocytometric data at multiple levels of tissue organization, ranging from individual cells and niches to atlas-defined brain regions\textsuperscript{4,13}. Although this pipeline-based image analysis strategy is powerful and comprehensive, it is rigid and requires programming to address emergent needs.

Here, we present an alternative query-driven strategy for multiplex brain tissue IHC image analysis, based on a multiplex Visual Search Engine (mViSE)\textsuperscript{14,15} (Figure 1). It directly serves the core need in most image analysis tasks–identifying one or more cell populations, multi-cellular units, or tissue regions of interest based on their morphological and spatio-molecular characteristics and profiling them via expression maps. Our mViSE operates in two phases. In the learning phase (Figure 2A), it learns brain cell morphologies, spatial protein expression patterns (chemoarchitecture), cell arrangements (cytoarchitecture), and wiring patterns (myeloarchitecture) in an integrated manner fully automatically in an unsupervised manner, without the need for human annotations or intervention. Importantly, our method provides a clear visual confirmation of successfully learning the facet of brain cytoarchitecture defined by each panel, avoiding the "black box" problem (Figure 2B).
\begin{figure}[ht]
\centering
% To include your figure, uncomment the line below and upload Figure_1.pdf
\includegraphics[width=\textwidth]{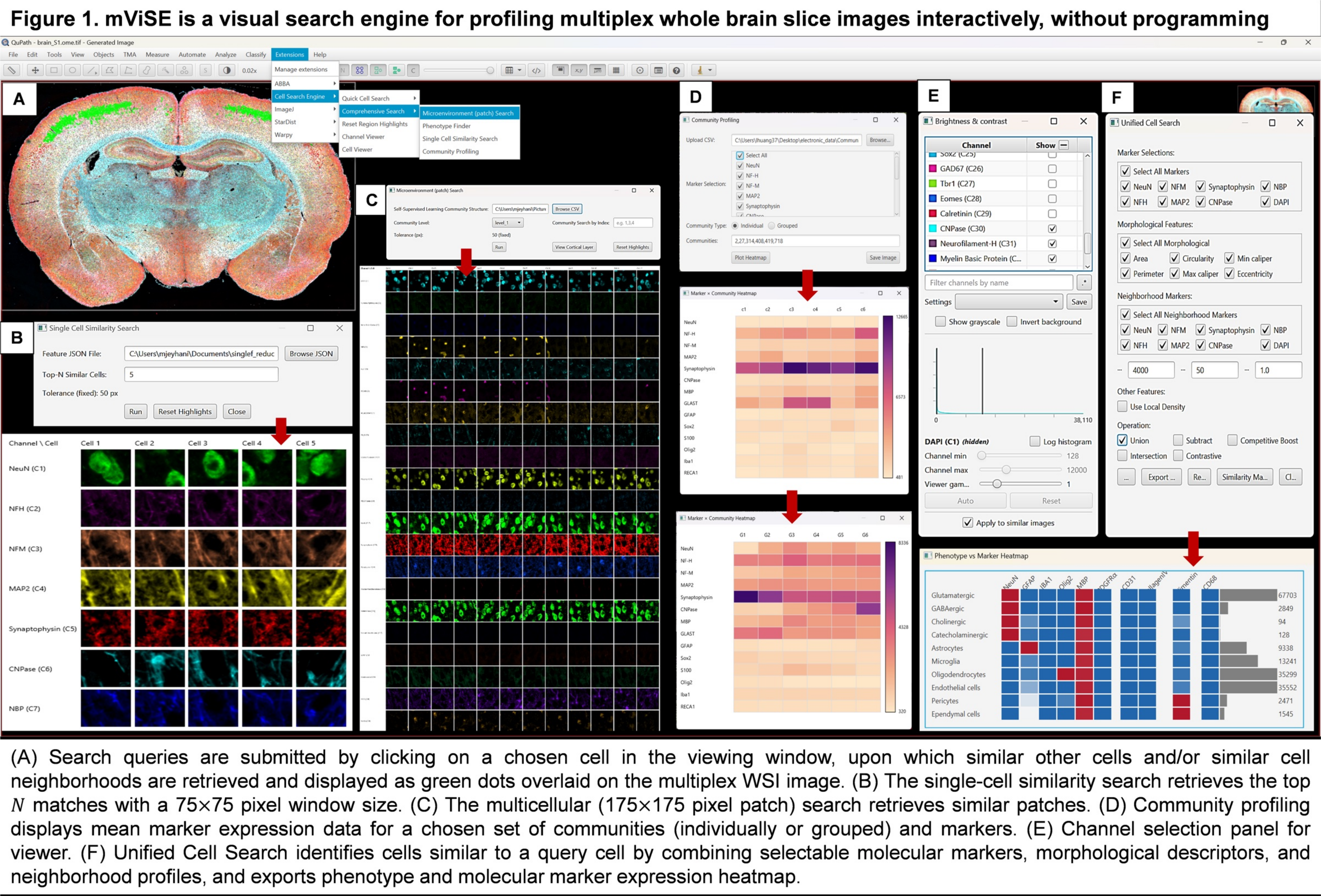}
\caption{mViSE system overview and workflow showing the two-phase operation: learning phase and query phase.}
\label{fig:fig1}
\end{figure}
Brain tissue has a multifaceted architecture. With this in mind, the learning process is conducted in a 'divide and conquer' manner. The multiplex data is organized into a group of panels of functionally related user-defined molecular markers (Figure 2C), one panel per facet. For example, a panel can consist of markers that describe the myeloarchitecture, from which mViSE learns the wiring patterns of brain tissue. Another panel can, for example, enable learning the distribution of neuronal subtypes, and so on. Collectively, these panels capture the multifaceted tissue architecture and can be used individually or in combination. In the query phase, the user can indicate cells or multicellular niches of interest, specify one or more relevant molecular panels, upon which mViSE retrieves similar communities of cells or niches, and displays their molecular profiles. This is made possible by a hierarchical strategy (Figure 2D) for combining the learned models for multiple panels. This is a direct, flexible, interactive, transparent, and versatile query-driven image analysis strategy that can be used for a variety of investigational goals, including tissue exploration, identification of cells and niches of interest, comparative profiling of cell populations and niches across brain regions, delineating brain regions and cell layers, regional profiling of molecular markers, and others.
\begin{figure}[ht]
\centering
% To include your figure, uncomment the line below and upload Figure_2.pdf
\includegraphics[width=\textwidth]{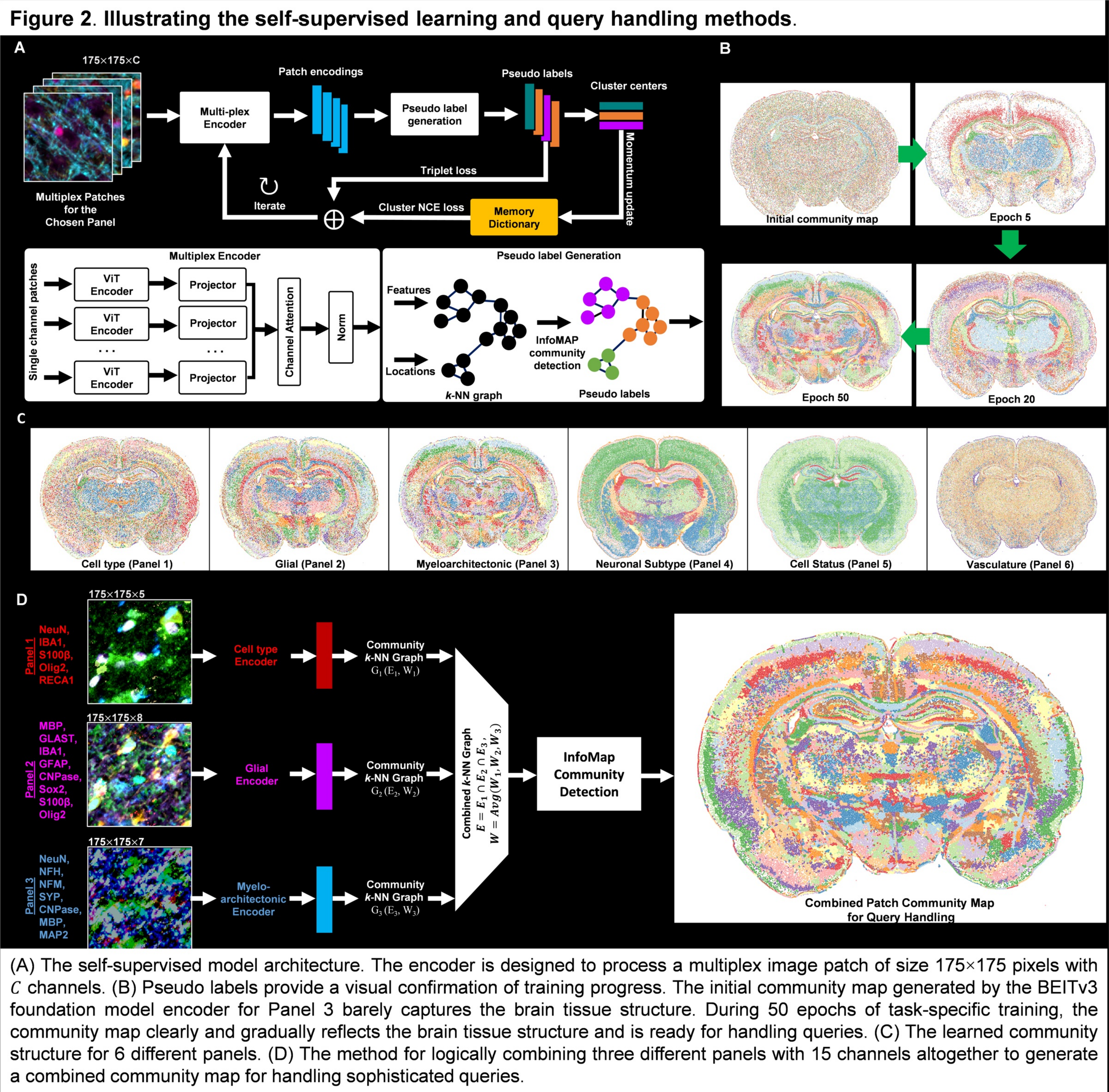}
\caption{Iterative training process illustrated in (A), showing the multiplex encoder training workflow with BEiTv3 initialization, channel attention module, and community detection. Visual confirmation of learning progress (B) through pseudo-label visualization. Panel combination strategy (C) demonstrating how multiple molecular marker panels are integrated. Successfully trained encoder results (D) showing pseudo-label maps capturing brain tissue architecture across six different panels.}
\label{fig:fig2}
\end{figure}
The design of mViSE for multiplex brain tissue images is not straightforward. Content-Based Medical Image Retrieval (CBMIR) systems that have been developed to identify medical images within large databases\textsuperscript{16-22} cannot handle high-order multiplex data. For example, Yottixel\textsuperscript{14}, a digital pathology search engine, represents whole-slide images as mosaics of patch barcodes, enabling real-time, large-scale retrieval with high accuracy and reduced storage requirements. UNI\textsuperscript{23}, a large-scale self-supervised pathology foundation model pretrained on a large WSI dataset across 20 tissue types, allows resolution-agnostic and data-efficient representation learning and outperforms prior models on several computational pathology tasks. Yet, existing applications work primarily for grayscale images or RGB color H\&E images; multiplex images remain largely unexplored, and the application of search engines to multiplex fluorescence imaging is still an open challenge.

Overcoming this challenge requires the design of multiplex encoders capable of capturing the complex and region-specific morphological and spatio-molecular patterns and their variations at the level of individual cells, tissue patches, and brain regions convincingly. This is challenging for multiple reasons. First, deep encoders with sufficient representational capacity are needed, but they generate high-dimensional encodings. We expect the high-dimensional encodings for visually similar patches to be proximal in the high-dimensional feature space, but this property is not a given. We found that a generic encoder, e.g., the variational autoencoder (VAE)\textsuperscript{24,25}, the U-Net\textsuperscript{26} with channel attention\textsuperscript{27}, or even the recently introduced transformer based multi-task foundational model (BEiTv3)\textsuperscript{28} do not meet this expectation\textsuperscript{29} (Figure 3B). Achieving this property requires that the encoder is specifically trained for handling the visual search query-driven retrieval task of interest to us. For this reason, we adapted techniques from the field of person re-identification\textsuperscript{30} that enable such task-specific encoder training. Another challenge emanating from the high-dimensional nature of neural network encodings is their inherently "black box" nature. To overcome this, we developed methods that provide an explicit visual confirmation of the trained encoder's ability to capture the known brain cytoarchitecture (Figures 2B, 2C, 2D).
\begin{figure}[ht]
\centering
% To include your figure, uncomment the line below and upload Figure_3.pdf
\includegraphics[width=\textwidth]{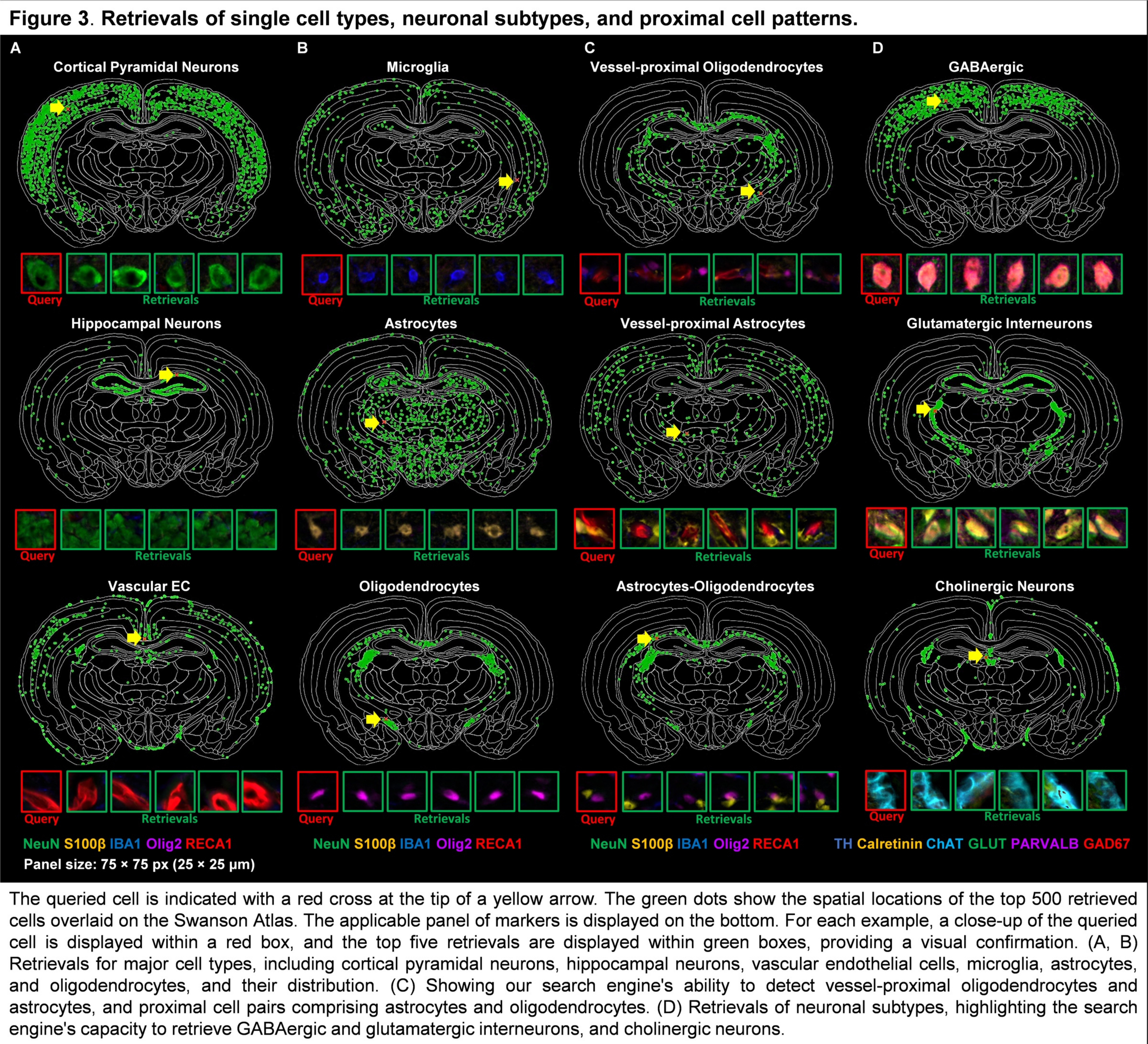}
\caption{Single cell retrieval examples showing query patches and the spatial distribution of the 500 most similar retrieved patches overlayed on the Swanson atlas. Retrievals shown for various cell types including cortical neurons, hippocampal neurons, RECA1+ vascular cells, IBA1+ microglia, S100+ astrocytes, Olig2+ oligodendrocytes, vessel-proximal cells, GABAergic neurons, glutamatergic interneurons, and cholinergic neurons. Top five retrievals are shown as close-ups for visual confirmation.}
\label{fig:fig3}
\end{figure}
For modeling the complexity of brain tissue, an encoder must have sufficient representational capacity to capture long-range spatio-molecular relationships and integrate information across multiple imaging channels in a meaningful manner. For this, we first adopted a Vision Transformer (ViT)-based encoder to provide sufficient representational capacity for multiplex imaging data across extended spatial regions, better than conventional CNN-based models such as U-Net or VAE\textsuperscript{25}. We also found that VAEs have an unavoidable built-in Gaussian smoothness prior that limits the model's ability to detect subtle differences across region boundaries in feature space\textsuperscript{31}. Vision Transformer models have the capacity to model details, complex long-range relationships, and subtle differences\textsuperscript{32}. For handling brain tissue, we found that the multivariate distribution of encodings is inherently unbalanced since some brain regions are much smaller and statistically under-represented compared to much larger regions, leading to under-fitting of the rare samples. The VAEs suffer from this limitation, in addition to bias\textsuperscript{33,34}. To overcome this limitation, we developed an iterative strategy combining $k$-nearest neighbor graphs and an information-theoretic community detection algorithm (InfoMap)\textsuperscript{35} – both scalable algorithms that treat all samples equally, and ensuring that small communities are preserved (Figure 2A). Furthermore, our training process integrates a contrastive loss and a triplet loss function to explicitly target discriminative representation learning. It prioritizes maximizing inter-class separability while minimizing intra-class variability, thereby enhancing the handling of imbalanced data, mitigating noise from irrelevant channels, and maintaining scalability to large datasets. Importantly, our process does not require parameter settings or tuning.

Each of the MP-WSI image channels represent specific proteins that have specific biological functions individually, and in combination with other proteins that may be present or absent. Unfortunately, current encoders are designed for 3-channel color RGB images and cannot handle MP-WSI images containing dozens of image channels requiring combinatoric interpretation. Merely extending current encoders (e.g., VAE) with additional channels becomes ineffective for 5 or more channels due to noise accumulation and instability in reconstructing multiplex image data. A more meaningful and a scalable strategy is clearly needed for handling multiplex images with more channels. In this work, we developed a "divide and conquer" strategy in which the image channels are organized into a set of panels, each containing a list of meaningfully related protein markers, and mViSE is provided with clear guidance on the combination of panels that are relevant for a given query, and a simple logical way to combine information across panels.

Below, we describe our multifaceted strategy to overcome the above-mentioned limitations, and a usable open-source software implementation in the form of a plug-in tool incorporated into the Qu-Path software that is widely used to analyze multiplex WSI images.

\section{Methods}
The detailed specimen preparation and imaging protocols, and the MP-IHC whole slide images (WSI) used here are disseminated in our earlier paper\textsuperscript{1} so not repeated here.

\subsection{Multiplex Visual Search Engine Design}
mViSE is trained on tissue patches over the set of images to be analyzed, or a representative subset. We used two patch sizes. Small patches (25 $\times$ 25 pixels, 330nm/pixel) centered on cell nuclei include individual cell bodies, perhaps a few basal processes, or a neighbor. Larger patches (175 $\times$ 175 pixels) include multicellular neighborhoods with multiple cell types, local myeloarchitecture, vascular cells or capillary fragments, etc. Other patch sizes can be used, provided the same size is used for training and retrieval. Although not essential, we find it advantageous to center the larger patches on individual cells and we refer to them as cell-centered patches (CCP). This is helpful when we are interested in the neighborhoods of specific cell types of interest. CCPs require prior cell detection, but the methods are now well established\textsuperscript{36}.

\subsection{Selection of Molecular Marker Panels}
Since multiplex imaging data contains a mix of channels, some of which are relevant, while other are irrelevant to an intended query, we found it best to allow the user to specify the panel of relevant molecular markers. For example, glial status markers would be irrelevant for a user interested in exploring neuronal arrangements. When the multiplexing is of high order (>10), we found it advantageous to organize the data into a set of panels of molecular markers. These panels are not set in stone – a user can define multiple customized panels and use them in combination as described below. We found panels consisting of 5 – 8 markers to work best. Larger panels suffer from imaging noise accumulation in the multiplex encoding steps described next. For the examples presented here, we defined six panels of molecular markers to capture various facets of brain tissue, including a cell-type panel \{NeuN, S100, IBA1, Olig2, RECA1\}, a glial panel \{S100, Sox2, GFAP, GLAST, Olig2, CNPase, MBP, IBA1\}, a myeloarchitectonic panel \{NeuN, NF-H, NF-M, MAP2, Synaptophysin (SYP), CNPase, MBP\}, a neuronal subtypes panel \{TH, Calretinin, ChAT, GLUT, PARVALB, and GAD67\}, cell phenotypic status panel \{DAPI, Histones, PCNA, CC-3, Nestin\}, and a vascular panel \{RECA1, Tomato Lectin, GFP, Blood Brain Barrier, Smooth Muscle Actin\}.

\subsection{Multiplex Vision Transformer Panel Encoders}
We trained an array of multiplex tissue patch encoders with weight vectors for each of the marker panels defined above, that accepts a tissue patch as input and computes an encoding vector, with the expectation that a pair of visually similar patches have a shorter distance compared to a pair of dissimilar patches\textsuperscript{27}, without the need for human labeling. For retrieving patches that are cytologically and/or cytoarchitecturally similar, and organizing the retrievals based on their similarity, the encoder must extract usable semantic and positional features of the relevant tissue constituents across the image channels.

The design of the encoder is crucial since it defines the effectiveness of mViSE. As noted above, a generic encoder, e.g., the variational autoencoder\textsuperscript{24} does not guarantee the distance requirement, fails to capture long-range relationships, and lacks representational capacity for multiplex image data. To overcome these limitations, we selected and iteratively trained the array of encoders with a cluster-contrastive learning loss function\textsuperscript{37} in the context of a retrieval task of specific interest to us, inspired by prior work in unsupervised person re-identification\textsuperscript{38-41}. In these systems, clustering algorithms are used to generate pseudo-labels that provide a robust supervisory signal for optimizing the encoder's parameters via backpropagation. Unfortunately, a direct application of re-identification methods is not effective for tissue similarity search in multiplex images since they are designed for Red Green Blue (RGB) images, the query patch simply does not recur, and the natural biological variability across the tissue is greater than the inter-camera variations encountered in person re-identification systems. On the other hand, brain tissue exhibits characteristic patterns of cell morphology, cellular arrangements, and characteristic expression patterns of molecular markers over brain regions, suggesting that tissue patches may be amenable to clustering (grouping), although the clustering criteria can vary greatly.

We initialized the iterative encoder training with the Bidirectional Encoder representation from Image Transformers (BEiTv3)\textsuperscript{28}, a large foundational model with a strong ability to capture morphology and positional patterns, detect long-range dependencies across an image, capture complex spatial correlations with low inductive bias, better multi-task versatility and adaptability and robustness compared to CNN-based encoders\textsuperscript{42}. A single ViT\textsuperscript{43} encoder with shared weights is applied independently to each channel, enabling consistent feature extraction while reducing model complexity (Figure 2A). However, as BEiTv3 is pre-trained on RGB images, rather than 5–8 channel multiplex data, we used an adaptive Efficient Channel Attention (ECA)\textsuperscript{44} module to dynamically emphasize the most informative channels.

Even the powerful BEITv3 encoder with the ECA is only weakly successful in capturing the brain cytoarchitecture as evident in the initial community map shown in Figure 2B that weakly reflects the brain tissue architecture. As noted earlier, this is because the encoder lacks task-specific training. Our iterative training process illustrated in Figure 2A that is inspired by a re-identification task fills this gap by minimizing a custom two-part loss function composed of a contrastive loss term and the triplet loss term as described below. For this, we used the Cluster Noise Contrastive Estimation (ClusterNCE) loss proposed by Dai et al.\textsuperscript{30} We model the known repetitive brain tissue architecture by organizing the patch data into clusters whose centroids in feature space form memory centroids. We refer to the list of cluster centroids as the memory dictionary (Figure 2A), and cluster indices as the pseudo labels\textsuperscript{30}. With this notation, for a point $q$ in feature space, the loss is given by:

\begin{equation}
L(q) = -\log \left( \frac{\exp(q \cdot c^+ / \tau)}{\sum_{k=1}^{K} \exp(q \cdot c_k / \tau)} \right)
\end{equation}

where $c^+$ is the center of the positive cluster that $q$ belongs to, $c_k$ is the center of the $k$-th cluster, "$\cdot$" indicates a dot product, and $\tau$ is a temperature term that controls the sensitivity of the loss function. We found that the cluster-contrastive loss alone fails to discriminate well in brain regions with low variability. To address this, we incorporated the triplet loss\textsuperscript{45}, that fosters the learning of a discriminative feature embedding space by minimizing the distances between similar patches and maximizing the distance between dissimilar patches. It is written in terms of the hardest positive (similar) sample and the hardest negative (dissimilar) sample, separated by at least a margin $m$ as follows:

\begin{equation}
L = \max \left( d(a,p) - d(a,n) + m, 0 \right),
\end{equation}

where $d$ is the Euclidean distance, $a$ is the anchor, $p$ is the positive sample, and $n$ is the negative sample. We found that when training with contrastive loss alone, the network consistently converged on two main clusters: unmyelinated and myelinated tissue, leaving it a suboptimal choice for learning similarities between other tissue types. Training with triplet loss alone produced weak associations across all tissues, and the delineation from one tissue type to another, specifically amongst cortical layers, was not sufficiently clear. However, the combination of these loss terms proved effective for mViSE, especially in discriminating regions with subtly different cytoarchitectures, e.g., the brain cortical layers.

The pseudo labels noted above are needed to compute the loss function, and they provide the supervisory signal for optimizing the encoders within our self-supervised training framework. At each iteration, we use the current draft encoder to generate encodings for each patch. From these encodings, we first construct a weighted undirected $k$-nearest neighbor graph\textsuperscript{46} in which the nodes are patches and the links are established for the most proximal nodes as defined by the following proximity measure between patches $i$ and $j$:

\begin{equation}
s(i,j) = \text{sim}(i,j) \exp \left( - \text{dist}(i,j) / \sigma \right),
\end{equation}

where $\text{sim}$ is the cosine similarity measure, $\text{dist}$ is the spatial distance between the patch centers, and $\sigma$ is a fixed weight parameter designed to add a modest spatial proximity bias. Each link is weighted by the corresponding similarity. The $k$-nearest neighbor graph forms the input to the two-level InfoMap algorithm\textsuperscript{35}. This generates a map of clusters (communities) of patches such that the information theoretic distance, measured as a code length is minimal for patches within the same cluster, and these are the pseudo labels. From the pseudo labels, we update the cluster centers using the following momentum update strategy needed to ensure the consistency of pseudo labels from one iteration to the next:

\begin{equation}
c_k \leftarrow (1 - \mu) c_k + \mu \bar{c_k},
\end{equation}

where $\mu$ is the momentum coefficient and $\bar{c_k}$ is the mean of features in cluster $k$. The updated cluster centers are used to compute the loss for performing a backpropagation update of the encoders. Pseudo labels can be displayed as color coded images (Figure 2), providing a visual confirmation of the multiplex encoder's progress during training as iterations (epochs) progress (Figure 2B). A successfully trained encoder reflects the brain tissue architecture implied by the specified panel of molecular markers when we visualize the pseudo labels (Figure 2D). An encoder is thus trained for each panel of chosen molecular markers.

\subsection{Query Processing: Single Cell Retrieval}
Once trained, the multiplex encoders transform our patches into 4,608-dimensional vectors that are used for processing queries, and we developed two strategies for this. To enable fast retrievals, we use principal components analysis to reduce the dimensionality of these vectors to 128. When a user indicates a query patch, the first strategy simply computes the cosine similarity with every other patch, and retrieves the top $N$ most similar patches, where $N$ is a number set by the user, e.g., 500 (Figure 2). We found this strategy to be useful for processing small patches (25 $\times$ 25) that typically contain a single cell. The most appropriate panel of molecular markers for the encoder for single-cell retrieval consists mostly of cell-type markers.

\subsection{Query Processing: Similar Multi-cellular Patch Retrieval}
For handling queries representing larger multi-cellular patches (175 $\times$ 175), we found it best to retrieve and display the community of patches of which the query is a member. The user can run this query using any of the trained panel-specific encoders (Figure 2D). This strategy is effective when working with panels of 5 – 8 molecular markers at a time.

A more sophisticated strategy is appropriate when a larger number of molecular markers, and therefore, multiple panels of molecular markers may be relevant to the query goal. The process for handling this convergence of information is illustrated in Figure 2C in the context of 3 panels (Cell type, Glial, and Myelo-architectonic panels). The encodings for each panel are used to form a $k$-nearest neighbor graph of all patches. These panel-specific graphs are combined into a single graph by including only the graph links that are common to all the graphs, i.e., a set intersection operation. The weights of the graph links are average of the weights. We then run the InfoMap algorithm to identify the resulting communities of patches. As evident in Figure 2C, the resulting community map intelligently combines information across the panels. In this case, when a user indicates a query patch, we simply retrieve the community of patches which the query is a member from the combined community map.

\section{Experimental Results}
We evaluated mViSE on the MP-IHC whole slide images (WSI) disseminated in our earlier paper\textsuperscript{1}. Note that some of the molecular markers can be present in more than one panel, and this does not pose a limitation for mViSE. We trained an encoder for each of these panels. Figure 2D provides a visual confirmation of our pseudo-label maps capturing the underlying brain tissue architecture over the six panels.

For single-cell retrieval experiments with 25 $\times$ 25 image patches, we used the cell-type panel, and the neuronal subtype panel. Figure 3 illustrates the retrieval of 25 $\times$ 25 image patches starting from a single query (indicated in red) and retrieving the 500 most similar patches/cells whose spatial distribution is indicated as green dots overlayed on the fitted Swanson atlas\textsuperscript{4} to provide a spatial reference for judging the retrievals of similar cortical neurons, hippocampal neurons, RECA1+ vascular cells, IBA1+ microglia, S100+ astrocytes, Olig2+ oligodendrocytes, vessel-proximal oligodendrocytes, vessel-proximal astrocytes, patches containing astrocytes and oligodendrocytes in close proximity, GABAergic neurons, Glutamatergic interneurons, and cholinergic neurons. For each query, the top five retrievals are presented as close-ups for visual confirmation. The full retrieval is provided in the electronic Supplement.

To validate the single-cell retrieval performance, we computed the top-$k$ retrieval scores using the results from our previous study as ground truth (Figure 5D). Our method achieved an average Top-1 accuracy of 0.90 and a Top-5 accuracy of 0.96 across the five major cell types. Notably, the RECA score, which marks endothelial cells, was comparatively lower due to vessel-proximal structures (Figure 3C). Beyond cell type classification, our approach also captures multiplex morphological features, enabling accurate retrieval in challenging scenarios where small, densely packed cells are overlapping.

Next, we evaluated mViSE for its ability to delineate small brain regions using a combination of three panels. Combining the panels resulted in a higher quality community structure as measured by the lower average code lengths, as summarized in Figure 5E.

The code length\textsuperscript{47}, is an information-theoretic metric used to evaluate the quality of a community structure. It quantifies the efficiency of a random walk constrained within the same community in the feature space. A lower code length indicates a better-defined community structure. Each panel's community code length reflects the quality of its corresponding structure. Panels 1, 2 and 3 community organization exhibits capacity to capture the global brain structure and delineate the functional regions, while Panels 4, 5 and 6 tend to capture the single cell level information (Figure 2D and Figure 6). Importantly, the combination of panels yields a lower code length, and the corresponding community structure effectively captures higher-order relationships, enabling more discriminative and robust community representations, which can also be observed visually in Figure 2C.
\begin{figure}[ht]
\centering
% To include your figure, uncomment the line below and upload Figure_6.pdf
\includegraphics[width=\textwidth]{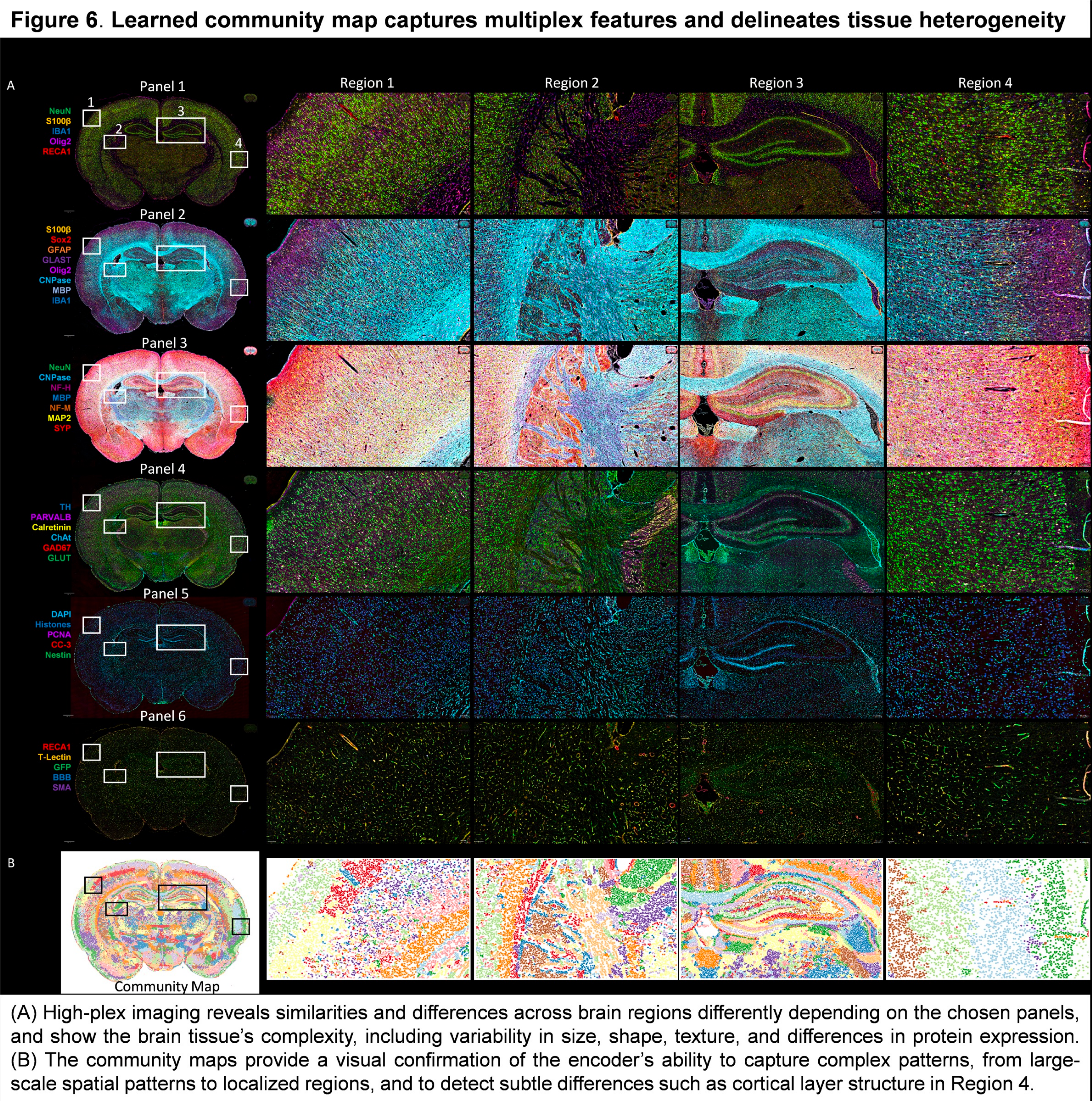}
\caption{Detailed microenvironment community analysis for four selected tissue regions (A, B, C, D). Bottom row shows combined community maps illustrating the model's ability to capture subtle differences in tissue architecture, including small, statistically under-represented regions. Examples from RSGC and Hippocampus demonstrate sensitivity to intra-regional architectural variations.}
\label{fig:fig6}
\end{figure}
Figure 4 illustrates the retrieval of similar 175 $\times$ 175 multi-cellular image patches, also starting from a single query, in the RSGc, mt, IGP, and LHB regions of the rat brain. The retrieved image patches are visually similar and delineate these brain regions effectively. The full retrieval is provided in the electronic Supplement.
\begin{figure}[ht]
\centering
% To include your figure, uncomment the line below and upload Figure_4.pdf
\includegraphics[width=\textwidth]{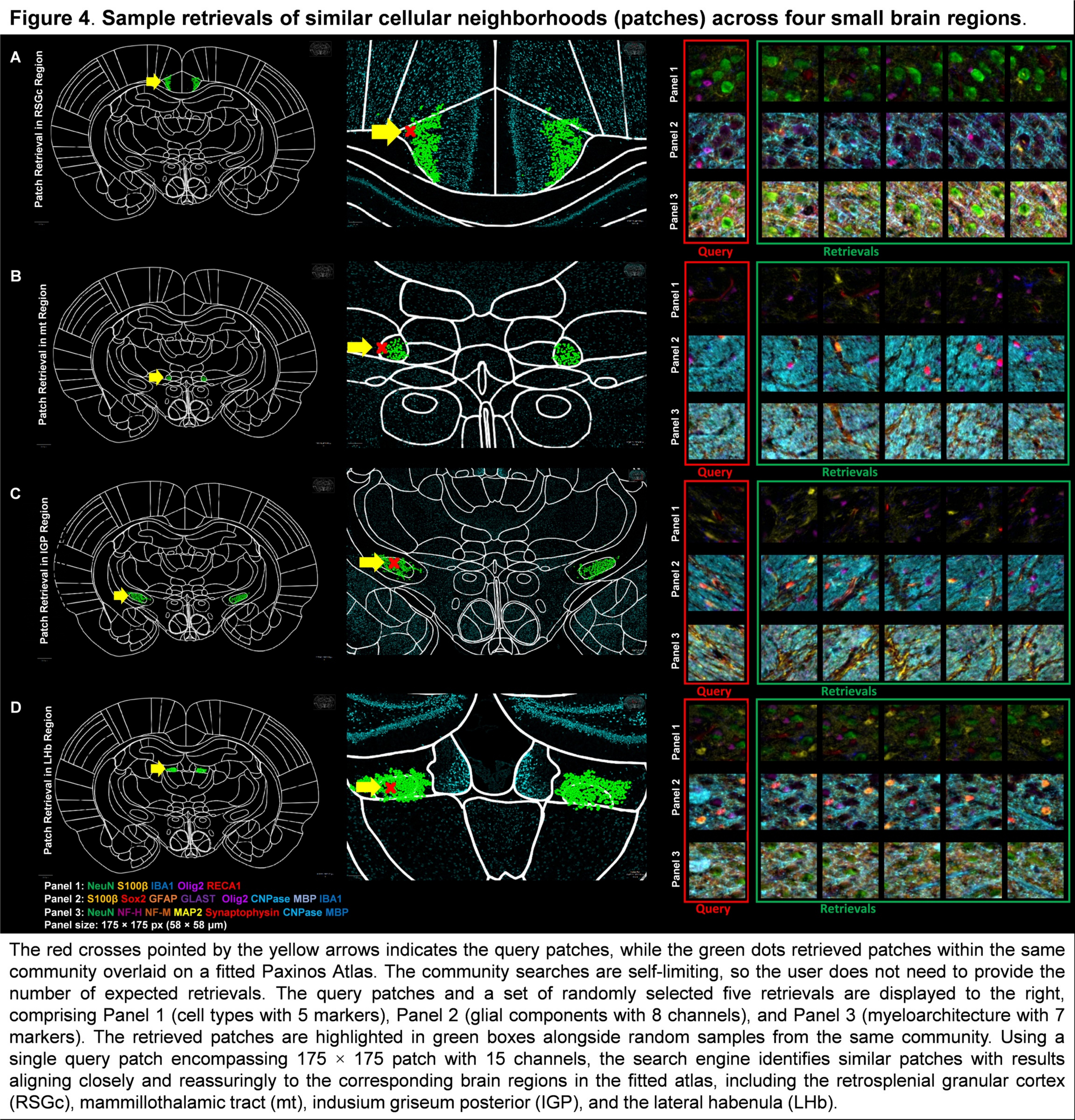}
\caption{Multi-cellular patch retrieval results (175 $\times$ 175 pixels) demonstrating the ability to retrieve similar tissue patches and effectively delineate brain regions including RSGc, mt, IGP, and LHB regions of the rat brain. Retrieved patches show visual similarity and consistent cytoarchitectural patterns.}
\label{fig:fig4}
\end{figure}
Using the same three panels, we evaluated the ability of mViSE to effectively delineate the cortical cell layers, a challenging task, generate molecular profiles of the layers, and further, to reveal the inter- and intra-layer variations. Figure 5 shows the results for Layers L1 - L6. Each cortical layer consists of a union of multiple communities in the combined community map. The first two rows of Figure 5 show the individual communities for each layer in different colors. The third row shows the expression maps for each of the layers for each panel. The 4th and 5th rows provide additional details for L1. The fourth row of Figure 5 shows the representative patches for L1, under each of the three panels. The 5th row of Figure 5 shows the corresponding expression maps.

To assess the performance of mViSE, we manually fit the Swanson Atlas\textsuperscript{4} to the images and calculated the confusion matrix and Intersection over Union (IoU) scores (Figure 5D). Our method achieved true positive rates of 0.8–0.9 across distinct cortical layers, with an average IoU of 0.70. These results indicate a high level of accuracy in capturing cortical organization in a fully unsupervised manner from the multiplex data, without customized programming, driven only by visual queries. This highlights the versatility and practical usefulness of mViSE. Figure 6 provides additional close-up illustrations of microenvironment communities for four selected tissue regions \{A, B, C, D\}. The bottom row of Figure 6 shows the combined community maps for panels illustrating the ability of the model to capture subtle differences in tissue architecture, including small regions that are statistically under-represented. Figure 6 shows the ability of the model to sense subtle differences in tissue architecture within brain regions (RSGC, Hippocampus).
\begin{figure}[ht]
\centering
% To include your figure, uncomment the line below and upload Figure_5.pdf
\includegraphics[width=\textwidth]{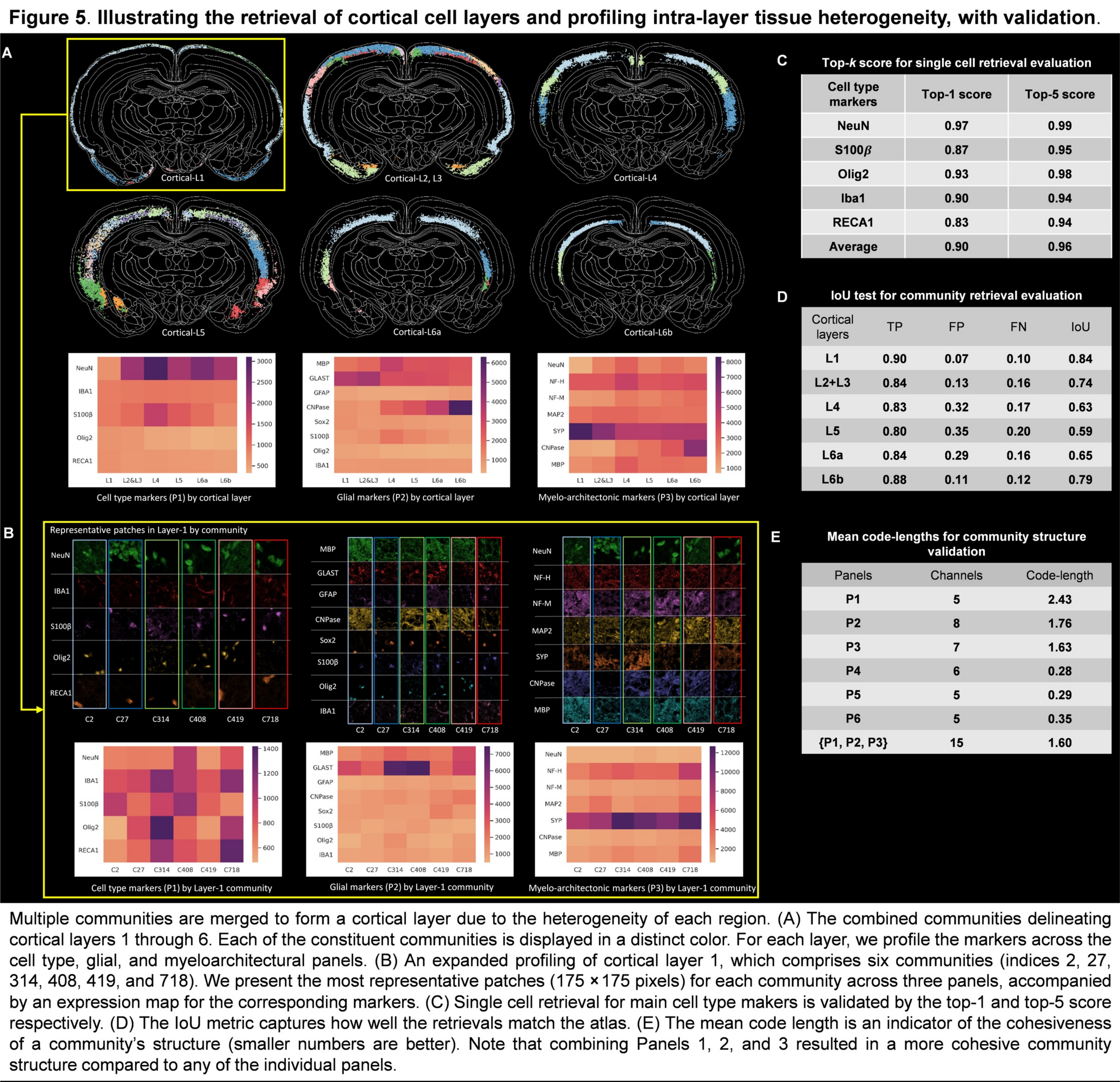}
\caption{Cortical layer delineation results for layers L1-L6. First two rows show individual communities for each layer in different colors. Third row displays expression maps for each layer across different panels. Fourth and fifth rows provide detailed analysis of Layer L1, showing representative patches and corresponding expression maps. Performance metrics (D) show confusion matrix and IoU scores with true positive rates of 0.8-0.9 and average IoU of 0.70. Code length comparison (E) demonstrates improved community structure quality when combining multiple panels.}
\label{fig:fig5}
\end{figure}
\section{Experimental Setup}
Dataset preparation—for a 16-bit, 30-channel image of 43,054 $\times$ 29,398 pixels containing 219,616 detected cells—took about 1.5 hours. Half of the datasets were used for training due to a dense cell-centered sample generation strategy. The encoder was initialized from the pretrained BEiTv3-base vision encoder, containing 86M parameters. To speed up training, the first nine layers of the encoder were frozen, leaving the final three layers trainable, combined with a linear projection head and the channel attention. Model training was parallelized on two NVIDIA RTX A6000 GPUs (48 GB) with a batch size of 256 for each panel. For a 5-channel panel with 109,808 samples and a patch size of 175 $\times$ 175 $\times$ 5 pixels, training converged in 6 hours, and computation of the community structure took 7 minutes. In practice, it is best to perform the model training on a GPU cluster overnight. The results of the training are saved as a community index table (in the .csv file format) containing the community index for each cell and for each panel, and the intensities of each of the protein marker signals. This table is used to handle the user's visual queries as described below.

We developed an extension for the widely used QuPath\textsuperscript{48} viewer to display the multiplex data, accept visual queries from the user, and display the retrievals, along with expression maps for the retrieved cells or patches. A query is handled simply by displaying the members of the community that the queried cell belongs to. This simple strategy is surprisingly effective since the retrieved items are optimal in the sense of being most proximal as measured by the information-theoretic code length. It is also very fast (sub-second). As our results show, most retrievals also delineate a brain region or a sub-region thereof, and this is a re-assuring visual confirmation of correctness. We provide a comprehensive tutorial covering the machine learning model, and the QuPath extension on our GitHub repository.

Recognizing that many users may not have access to a GPU cluster, we also developed a "quick search" tool that does not utilize the neural network model and can be run fully within QuPath. This tool can also be used along-side the deep model based (comprehensive) search engine. Specifically, the user can use QuPath tools to identify all cells, compute cell morphometrics, establish cell phenotypes, and compute the distribution of protein markers within an image patch centered on each cell. As illustrated in Figure 1, the user can select a query cell and flexibly choose among molecular markers, morphological features, and neighborhood cell markers. The system then returns the top most similar matches (as specified by the user) along with the corresponding similarity matrix.

\section{Conclusions and Discussion}
The multiplex visual search engine offers a powerful, intuitive, logical, scalable, interactive, unsupervised, versatile, and a broadly applicable strategy for analyzing high-order multiplexed images without the need for complex programming, and the QuPath implementation makes it widely accessible to the scientific community. Importantly, the model training is fully automatic and does not require any human annotation efforts. Importantly also, our training process culminates with a clear visual confirmation of successfully capturing the facet of tissue architecture described by each panel. At the next level, our strategy allows the user to specify the combination of panels to be used to process a query, and this provides a degree of transparent control over mViSE.

mViSE enables a query-driven style of investigation of multiple facets of cellular and tissue architecture, assisting with diverse studies in basic and applied neuroscience. The model seamlessly integrates morphological and molecular information in a logical manner since the panels can be defined by the user to reflect the properties of the molecular markers. As noted above, query retrievals are easy to interpret since they often delineate known brain regions or sub-regions, and the accompanying expression maps reflect the protein distributions for the retrieved community of cells or cellular neighborhoods. This made it possible for us to quantify the quality of retrievals by fitting a brain atlas to the image and measuring how well the relevant brain region / sub-region was delineated by the retrievals.

In this paper, we focused on healthy brain tissue in our examples given the ease of visual confirmation, but the proposed method is equally applicable to diseased or injured brains. These data will be presented separately, along with the applicable differential profiling methods. As the number of imaging channels grows with instrumentation advances, we expect the search engine based image analysis strategy to grow to be especially valuable due to its inherent simplicity and versatility. In future work, we see opportunities to accelerate the system using model quantization and adoption of faster GPU systems to cope with hyper-plex imaging data on a large scale. We also foresee opportunities to strengthen the search engine for direct comparative profiling of brain regions across tissue samples.

\bibliographystyle{unsrtnat}

\clearpage

\end{document}